\documentclass[twocolumn,showpacs,preprintnumbers]{revtex4}

\usepackage{amssymb}
\usepackage{amsmath}
\usepackage{graphicx}
\usepackage{dcolumn}
\usepackage{bm}

\begin{document}

\title{Retrieval of phase memory in two independent atomic ensembles by Raman process}
\author{Cheng-ling Bian$^{1}$}
\author{Li-Qing Chen$^{1,3}$}
\author{Guo-Wan Zhang$^{1}$}
\author{Z. Y. Ou$^{1,2*}$}
\author{Weiping Zhang$^{1,3\dag}$}
\affiliation{$^{1}$Quantum Institute for Light and Atoms, Department of Physics, East China Normal University, Shanghai 200062, P. R. China}
\affiliation{$^{2}$Department of Physics, Indiana University-Purdue University Indianapolis,
402 N. Blackford Street, Indianapolis, IN 46202, USA}
\affiliation{$^{3}$State Key Laboratory of Precision Spectroscopy, East China Normal University, Shanghai 200062, P. R. China}

\date{\today }

\begin{abstract}
In spontaneous Raman process in atomic cell at high gain, both the Stokes field and the accompanying  collective atomic excitation (atomic spin wave) are coherent. We find that, due to the spontaneous nature of the process, the phases of the Stokes field and the atomic spin wave change randomly from one realization to another but are anti-correlated.  The phases of the atomic ensembles are read out via another Raman process at a later time, thus realizing phase memory in atoms. The observation of phase correlation between the Stokes field and the collective atomic excitations is an important step towards macroscopic EPR-type entanglement of continuous variables between light and atoms.

\end{abstract}

\maketitle

Correlations in a quantum system played an important role in the test of foundation of quantum mechanics \cite{asp,ou92}, and in the applications of quantum metrology \cite{sci} and quantum information \cite{zei}. Recently, atomic memory for correlated photons is demonstrated \cite{luk1,kim} based on the DLCZ scheme \cite{DLCZ} in a collective Raman process in spontaneous emission regime for the application in quantum repeaters for long distance quantum communication.

But what is demonstrated so far is the intensity correlation between the Stokes field and the atomic excitations in the spontaneous Raman process. The phase correlation between the related optical fields and atomic medium  has not been explored.
With photon correlation demonstrated \cite{luk1,kim,kim05}, we believe that there is a strong correlation in the phases of the atomic ensemble and the Stokes field as well.

In this paper, we report on an experiment in which
we measure the phase difference of two spatially separated atomic spin waves created via spontaneous Raman processes in high gain regime and directly confirm a phase anti-correlation between the Stokes field and the corresponding atomic spin wave in the collective Raman process. The phase of the atomic spin wave is retrieved after a time delay of the phase measurement of the Stokes field, thus realizing the memory of phase information of optical fields.

\begin{figure}[tbp]
\centerline{\includegraphics[scale=0.45]{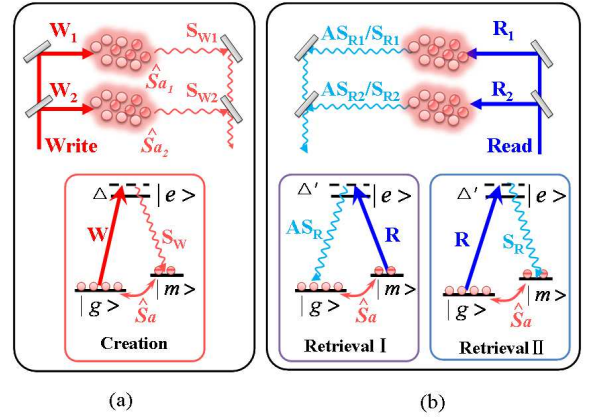}}
\caption{The conceptual diagrams for measuring phase correlation: (a) writing process;  (b) reading process with two methods (Retrieval I,II). Inset: atomic levels for Raman scattering.} \label{fig1}
\end{figure}

The conceptual sketch of the experiment is shown in Fig.1. The basic process is a collective Raman scattering process in atoms with a lambda structure: an excited state $|e\rangle$ and two  meta-stable lower level states $|g\rangle,|m\rangle$ (insets of Fig.1). $N$ atoms are initially prepared in the ground state $|g\rangle$ by optical pumping. There are two stages of operation, depicted in Figs.1a and 1b, respectively. In the first stage shown in Fig.1a, two Raman pump pulses ($W_1,W_2$) start the Raman scattering to produce random Stokes fields $(S_{W1}, S_{W2})$ and the corresponding atomic spin waves $\hat S_{a1,2} \equiv (1/\sqrt{N})\sum_i |g_{1,2}\rangle_i\langle m_{1,2}|$ between two lower states in two separate atomic ensembles. We superimpose the generated Stokes fields to measure their relative phase. The second stage reads out the atomic spin waves by another Raman process as shown in Fig.1b and the relative phase of the readout fields are measured by interference methods. We look for the correlation between the two measured phases.

The interaction Hamiltonian in the write process is given by \cite{DLCZ}
\begin{eqnarray}
\hat H_R = \eta A_W^* \hat a_{S_W}\hat S_a + h.c., \label{HR}
\end{eqnarray}
where we treat the write field as a strong classical field with an amplitude of $A_W$. $\hat a_{S_W}$ denotes the Stokes field generated in the write process and $\hat S_a$ is for the atomic spin wave.
The quantum evolution of the Stokes field and the atomic spin wave is given by \cite{raymer}
\begin{eqnarray}
&&\hat a_{S_W}(t) = \hat a_{S_W}(0) \cosh \zeta t + \hat S_a^{\dag} (0) \sinh \zeta t,\cr
&&\hat S_a(t) = \hat S_a(0) \cosh \zeta t + \hat a_{S_W}^{\dag} (0) \sinh \zeta t,\label{SW}
\end{eqnarray}
where $\zeta \propto |A_W|$. With vacuum input for the Stokes field and atoms in the ground state, we can calculate the correlation function:
\begin{eqnarray}
\langle \hat a_{S_W}(t)\hat S_a(t)\rangle= \cosh \zeta t  \sinh \zeta t,\label{aS}
\end{eqnarray}
or the normalized correlation function
\begin{eqnarray}
\gamma_{a_{S_W}S_a} = {\langle \hat a_{S_W}(t)\hat S_a(t)\rangle \over \sqrt{\langle \hat S_a^{\dag}(t)\hat S_a(t)\rangle\langle \hat a_{S_W}(t)\hat a_{S_W}^{\dag}(t)\rangle}} = 1\label{gaS}
\end{eqnarray}
Note that this quantity is different from the coherence function of $\langle \hat a_{S_W}^{\dag}(t)\hat S_a(t)\rangle$, which vanishes here. Eq.(\ref{gaS}) indicates that the phase $\varphi_{S_W}$ of the Stokes field and the phase $\varphi_{S_a}$ of the atomic spin wave are anti-correlated via $\varphi_{S_W}+\varphi_{S_a} =$ const.

To confirm the phase anti-correlation, we need to measure the phases of the Stokes field and the atomic spin wave. Since phase is a relative quantity, we use two Raman processes to compare the relevant phases. The measurement of the phase of the Stokes fields is done by superposing the two Stokes fields to observe interference pattern. This was first observed by Kuo $et~al.$ \cite{kuo} in a Raman process in hydrogen cells and more recently by Chen $et~al.$ in a Rb atomic cell \cite{chen2}.  The interference pattern in our case is a frequency beat \cite{chen2}, as shown in the first part of Fig.2 (blue curve). The phases of the Stokes fields can be extracted from the beat signal (see below for more details).

\begin{figure}[tbp]
\centerline{\includegraphics[scale=0.55]{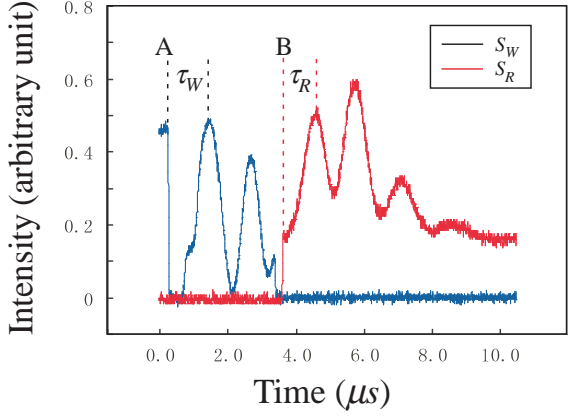}}
\caption{ Beat signals between Stokes fields for the write process (blue, first) and read process (red, second)  } \label{fig2}
\end{figure}

For the phase of the atomic spin waves, we need to read out the atomic spin waves and superpose the two readout fields (Fig.1b). The traditional way to read out the atomic coherence is through anti-Stokes field as in the well-known CARS technique \cite{shen}, shown as the $AS_R$ field in Fig.1b (Retrieval I). However, the CARS signal is typically very weak and short to exhibit any beat signal for phase measurement. Instead, we resort to an amplification feature in the same Raman process: when a seed is injected in the Raman process, it will be amplified, even though it may add in noise as spontaneous emission. Here, the seed is the atomic spin wave prepared in the first stage (Fig.1a). Recently, this amplification process was used to enhance Raman scattering \cite{chen09}. As in any three-wave mixing process, besides the amplification of the seed wave, there is an idler field generated. It has  about the equal size as the amplified field and carries the coherence of the seed. In the case here, the idler field is the Stokes field. So in the second stage of the experiment (Fig.1b), immediately after the first one, we send in reading pulses at similar frequency to the write lasers (Retrieval II). In order to preserve the coherence in the process, we split one laser into two reading beams. The generated Stokes fields $S_{R1}, S_{R2}$ are related to the initial atomic spin waves by \cite{raymer}:
\begin{eqnarray}
\hat a_{S_R}(t) = G \hat a_{S_R}(0)  + e^{i\varphi_R}\hat S_a^{\dag} (0) \sqrt{G^2-1},\label{S}
\end{eqnarray}
where $G$ is the amplitude gain of the amplifier. Then the coherence function between the two read Stokes fields is
\begin{eqnarray}
\langle \hat a^{\dag}_{S_{R1}}\hat a_{S_{R2}}\rangle  = (G^2-1)\langle \hat S_{a1} (0) \rangle \hat S_{a2}^{\dag}(0)\rangle e^{i(\varphi_{R2}-\varphi_{R1})}.\label{S12}
\end{eqnarray}
So the coherence of the two atomic spin waves is directly related to the coherence of the two Stokes fields generated by the reading fields. Here the initial Stokes field $\hat a_S(0)$ is in vacuum and $\varphi_R$ is the phase of the reading field.   We observe the interference between $S_{R1}, S_{R2}$ by mixing them. A typical interference pattern is shown as a frequency beat in the second part of Fig.2 (red line).
The beat signal in the reading process is delayed from the beat signal in the writing process by a fixed duration determined by the delay between the writing and reading pulses ($\tau \sim 0.1\mu s$  from the end of the write pulses. See the inset of Fig.2).

We may extract the phase information by measuring the locations ($\tau_W, \tau_R$) of the first peak in the beat signals relative to fixed reference points (A and B points in Fig.2): $\varphi_{S_R} = 360\times \tau_R/T_R, \varphi_{S_W} = 360\times \tau_W/T_W$. Here $T_W, T_R$ are the average periods of the beat signals for Stokes of write and read, respectively.

\begin{figure}
\centerline{\includegraphics[angle=-90,bb=0 0 550
800,scale=0.32]{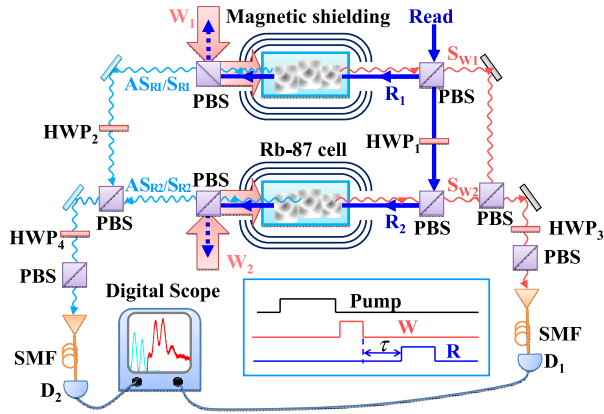}} \caption{(Color Online) Experimental
arrangement. PBS: polarization beam splitter; HWP: half wave plate
rotator; SMF: single-mode fiber; D: photo-detector. Inset:
timing sequence.} \label{fig3}
\end{figure}

A somewhat detailed experimental sketch is shown in Fig.3, together with timing sequence as inset. The two atomic ensembles are isotopically enriched Rb-87 without buffer gas, contained in two
cylindrical Pyrex cells (length and diameter are 75 mm and 19 mm,
respectively). The cells are respectively mounted inside three-layer magnetic shielding to reduce stray magnetic fields. The cells are heated
up to 72$^{\text{o}}$C using bifilar resistive heaters. Referring to the atomic energy levels shown in Fig.1, the two lower energy levels of the lambda structure are the hyperfine splitting of the ground states:  $|g, m\rangle = 5^{2}S_{1/2} (F=1, 2)$. The upper level is $|e\rangle = 5^{2}P_{1/2}$. The optical pumping pulses (Pump, not shown in experimental arrangement) are applied before the write pulses in order to initialize the atoms in the ground state $|g\rangle = 5^{2}S_{1/2} (F=1)$. The write pulses are from a single-frequency laser operating at 795 nm and are detuned from the Rb-87 $D_{1}$ line ($5^{2}S_{1/2}$ $F=1\rightarrow
5^{2}P_{1/2}$ $F^{\prime }=2$ transition) by $\Delta = 0.8 GHz$ for maximum Raman gain. The duration of the two write pulses is $3$ $\mu $s
and the powers are $11$ mW and $21$ mW, respectively. After a short delay of $\tau \sim 0.1\mu s$ from the end of the write pulses, we send in the reading pulses. They are derived from one laser by a polarization beam splitter (PBS) and have a duration of 16 $\mu s$. The powers of the reading beams ($R_{1,2}$) are 24 $mW$ and 3.2 $mW$, respectively. The generated Stokes fields are orthogonal to the reading fields in polarization and can be easily separated by PBS. The Stokes fields $S_{R1}$ is combined with $S_{R2}$ via a PBS and a half wave plate ($HWP_2$). Because of the different power in the reading beams, the generated Stokes fields also differ. We use a half wave plate ($HWP_2$) to adjust their relative intensities to obtain the maximum contrast in interference. The combined field is projected to the same polarization ($HWP_4$+PBS) before coupled into a single-mode fiber (SMF) for spatial mode clean-up and photo-detection. The Stokes fields from the write process undergo a similar arrangement. A digital scope monitors the temporal behavior of the detected signals with a typical result shown in Fig.2. The extracted phases from the two beat signals are shown in Fig.4a.

As observed in Refs.\cite{kuo,chen2}, the phases of the Stokes fields are random. Here we also observe  random phases for the atomic spin waves. The results of the phase measurement are shown in Fig.4a, where we plot $\varphi_{S_{R}}$ versus $\varphi_{S_{W}}$. A strong correlation of $\varphi_{S_{R}}-\varphi_{S_{W}} =$ const. is observed. This correlation is because both $\varphi_{S_{R}}$ and $\varphi_{S_{W}}$ are anti-correlated to the phase $\varphi_{S_{a}}$ of the atomic spin wave by $\varphi_{S_{R}} + \varphi_{S_{a}} = $ const. and $\varphi_{S_{W}} + \varphi_{S_{a}} = $ const. So Fig.4a is an indirect confirmation of the anti-correlation relation between $\varphi_{S_{W}}$ and $\varphi_{S_{a}}$.

\begin{figure}[tbp]
\centerline{\includegraphics[scale=0.35,angle=0]{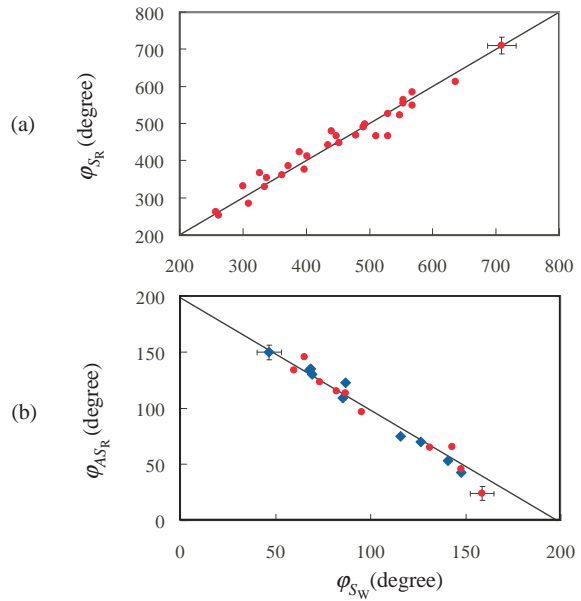}}
\caption{ (a) Phase of the read Stokes field $\varphi_{S_{R}}$ versus phase of the write Stokes field $\varphi_{S_{W}}$. The solid line has a slope of 1. (b) Phase of the read anti-Stokes field $\varphi_{AS_{R}}$ versus phase of the write Stokes field $\varphi_{S_{W}}$. The red circles and blue diamonds are two separate runs. The solid line has a slope of -1.}
\label{fig5}
\end{figure}

As discussed earlier and shown in Fig.1b, we can also retrieve the phases of the atomic spin waves by anti-Stokes (Retrieval I) process even though it is quite weak. But if the read field is strong enough, the anti-Stokes field can exactly represent the atomic spin wave $\hat S_a$ \cite{DLCZ,ou08}. For the anti-Stokes read process, we need to tune the read laser to $m\rightarrow e$ transition as shown in Fig.1b. The anti-Stokes process will then dominate over the Stokes process. From the interference signal $AS_{R_1}+AS_{R_2}$, we may likewise extract the phases $\varphi_{AS_R}$ of the anti-Stokes field, which is exactly the phase of the atomic spin waves.

The anti-Stokes field in the Raman process depends on the atomic spin wave as well as the strength of the reading field. It is usually weak and
has relatively short duration due to limited atomic excitation which results in a small atomic spin wave. Thus, it is very hard to observe multiple periods of the beat signal with the anti-Stokes fields. To extract the phases, we resort to the second method, where we measure simultaneously the individual and resultant intensities of the interfering fields. More specifically as shown in Fig.5, three detectors are used in this scheme where D1, D2 record the individual intensities $I_1, I_2$ of two interfering fields and D3 records the intensity $I_{12}$ of the superposed field: $i_1=\alpha_1 I_1, i_2=\alpha_2 I_2, i_3= \alpha_3 I_{12}$. Here $i_{1,2,3}$ are the photo-currents of D1, D2, D3, respectively and $\alpha_{1,2,3}$ are the corresponding efficiencies. A single mode fiber (SMF) is used for the superposition of the two anti-Stokes fields ($AS_{R1,2}$) in order to obtain good spatial overlap. A half wave plate (HWP) and polarization beam splitters (PBS) are used in order to adjust the relative strength of the two anti-Stokes fields. From the expression:
\begin{eqnarray}
I_{12} = I_1 + I_2 + 2\sqrt{I_1 I_2} \cos\varphi,\label{i}
\end{eqnarray}
where temporal dependence in the form of beat disappears when we set the beat frequency $\Delta \nu$ to zero by properly adjusting the powers of the reading fields, we can then extract the phase as
 \begin{eqnarray}
\cos\varphi &=& (I_{12} - I_1 - I_2) /2\sqrt{I_1 I_2} \cr
&=& {i_3 \sqrt{\alpha_1\alpha_2/\alpha_3^2} - i_2\sqrt{\alpha_1/\alpha_2} -i_1\sqrt{\alpha_2/\alpha_1}\over 2\sqrt{i_1i_2}}.\cr &&\label{ph}
\end{eqnarray}
The efficiency ratios $\alpha_1/\alpha_3, \alpha_2/\alpha_3$ can be calibrated by blocking one of the interfering fields.

\begin{figure}[tbp]
\centerline{\includegraphics[scale=0.33,bb=70 0 400
800,angle=-90]{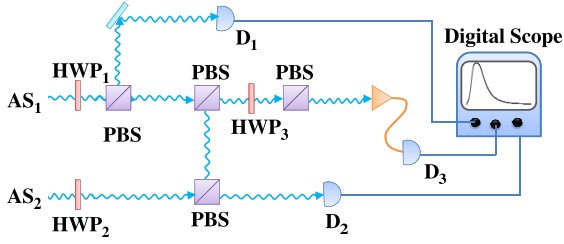}}
\caption{Schematics for measuring the phases of anti-Stokes fields.}
\label{fig7}
\end{figure}

To be consistent, a similar method is used to extract the phases of the Stokes fields at the same time as we extract the phases of the corresponding anti-Stokes fields.
In Fig.4b, we plot the phases of the read anti-Stokes field versus the phases of write Stokes field and obtain a strong anti-correlation $\varphi_{AS_R}+ \varphi_{S_W} =$ constant. This directly confirms the phase anti-correlation between the Stokes field and the atomic spin wave, which is the result of the phase conjugation of the two waves, as stated in Eq.(\ref{SW}).


The phase correlation revealed here between the Stokes field and the atomic spin wave together with the intensity correlation studied in photon correlation experiment \cite{DLCZ,kim,kim05} implies an amplitude correlation. In fact, this type of amplitude correlation is exactly the original EPR entanglement of quadrature phase amplitudes or continuous variables entanglement associated with any three-wave mixing process \cite{ou92}. In our case here, it is the entanglement of the quadrature-phase amplitudes of Stokes field and the atomic spin wave. Therefore, our experiment is an important step towards the demonstration of EPR-type continuous-variables entanglement between light and atomic ensemble. With two such systems, by making a projective measurement of continuous variables \cite{kim98} on the two Stokes fields, we should be able to put the atomic spin waves in the two atomic ensembles into an entangled state, thus realizing a quantum repeater for continuous variables.

This work is supported by the National Basic Research
Program of China (973 Program Grant No. 2011CB921604), the National Natural Science Foundation of China
(Grant Nos. 10828408, 10588402, and 11004058),  and the Program of
Shanghai Subject Chief Scientist (Grant No. 08XD14017).
\newline
Email:$^*$zou@iupui.edu; $^\dag$wpzhang@phy.ecnu.edu.cn

\end{document}